\documentclass[conference]{IEEEtran}

\usepackage{graphicx}

\usepackage[vlined, linesnumbered, ruled]{algorithm2e}

\let\labelindent\relax
\usepackage{enumitem}

\newcommand{\tuple}[1]{$\langle$#1$\rangle$}

\usepackage{color}

\newcommand{\commentmarkerfont}{\tiny}
\newcommand{\commenttextfont}{\scriptsize}

\newcounter{comment}
\newcommand{\comment}[2]{}

\newcommand{\makecommentmarker}[1]{%
  \begingroup
  \setlength{\fboxsep}{1pt}%
  \csname #1commentcolor\endcsname
  \fbox{\commentmarkerfont\thecomment}%
  \endgroup}
\newcommand{\commenttext}[2]{%
  \marginpar{%
    \raggedright
    \leavevmode
    \csname #1commentcolor\endcsname
    \makecommentmarker{#1}
    \commenttextfont #1: #2}}

\begin{document}


\title{QuPARA: \textbf{Qu}ery-Driven Large-Scale \textbf{P}ortfolio\\
  \textbf{A}ggregate \textbf{R}isk \textbf{A}nalysis on MapReduce}

\author{\IEEEauthorblockN{Andrew Rau-Chaplin, Blesson Varghese, Duane Wilson, Zhimin Yao, and Norbert Zeh}
  \IEEEauthorblockA{Risk Analytics Lab, Dalhousie University\\
    Halifax, Nova Scotia, Canada\\
    Email: \{arc, varghese, yao, nzeh\}@cs.dal.ca}}

\maketitle


\begin{abstract}
Modern insurance and reinsurance companies use stochastic
simulation techniques for portfolio risk analysis. Their risk portfolios may
consist of thousands of reinsurance contracts covering millions of individually
insured locations. To quantify risk and to help ensure capital adequacy, each
portfolio must be evaluated in up to a million simulation trials, each capturing
a different possible sequence of catastrophic events (e.g., earthquakes,
hurricanes, etc.) over the course of a contractual year.

In this paper, we explore the design of a flexible framework for
portfolio risk analysis that facilitates answering a rich variety of catastrophic
risk queries. Rather than aggregating simulation
data in order to produce a small set of high-level risk metrics efficiently (as
is often done in production risk management systems), the focus here is on
allowing the user to pose queries on unaggregated or partially aggregated
data. The goal is to provide a flexible framework that can be used by analysts
to answer a wide variety of unanticipated but natural ad hoc queries.
Such detailed queries can help actuaries or underwriters to better understand
the multiple dimensions (e.g., spatial correlation, seasonality, peril features,
construction features, financial terms, etc.) that can impact portfolio risk
and thus company solvency.

We implemented a prototype system, called QuPARA (Query-Driven Large-Scale
Portfolio Aggregate Risk Analysis), using Hadoop, which is Apache's implementation
of the MapReduce paradigm.
This allows the user to take advantage of large parallel compute servers in
order to answer ad hoc risk analysis queries efficiently even on very large
data sets typically encountered in practice.
We describe the design and implementation of QuPARA and present experimental
results that demonstrate its feasibility.
A full portfolio risk analysis run consisting of a 1,000,000 trial simulation,
with 1,000 events per trial, and 3,200 risk transfer contracts can be completed
on a 16-node Hadoop cluster in just over 20 minutes.
\end{abstract}

\begin{IEEEkeywords}
  ad hoc risk analytics; aggregate risk analytics; portfolio risk; MapReduce;
  Hadoop
\end{IEEEkeywords}

\section{Introduction}

\label{sec:introduction}

At the heart of the analytical pipeline of a modern
insurance/reinsurance company is a stochastic simulation technique for
portfolio risk analysis and pricing referred to as \emph{Aggregate Analysis}
\cite{s1, s2, s3, s4}.
At an industrial scale, a risk portfolio may consist of thousands
of annual reinsurance contracts covering millions of individually insured
locations. To quantify annual portfolio risk, each portfolio must be evaluated
in up to a million simulation trials, each consisting of a sequence
of possibiliy thousands of catastrophic events, such as earthquakes, hurricanes
or floods. Each trial captures one scenario how globally distributed
catastrophic events may unfold in a year. 

Aggregate analysis is computationally intensive as well as data-intensive.
Production analytical pipelines exploit parallelism in aggregate risk analysis
and ruthlessly aggregate results. The results obtained from production
pipelines summarize risk in terms of a small set of standard portfolio metrics
 that are key to regulatory bodies, rating
agencies, and an organisation's risk management team,
such as Probable Maximum Loss (PML) \cite{PML-1, PML-2} and Tail Value-at-Risk
(TVaR) \cite{TVAR-1, TVAR-2}.
While production
pipelines can efficiently aggregate terabytes of simulation results into a
small set of key portfolio risk metrics, they are typically very poor at
answering the types of ad hoc queries that can help actuaries or underwriters
to better understand the multiple dimensions of risk that can impact a portfolio,
such as spatial
correlation, seasonality, peril features, construction features, and financial
terms.

This paper proposes a framework for aggregate risk analysis that facilitates
answering a rich variety of ad hoc queries in a timely manner.
A key characteristic of the proposed framework is that it is
designed to allow users with extensive mathematical and statistical skills but
perhaps limited programming background, such as risk analysts, to pose a rich
variety of complex risk queries. The user formulates their query by defining
SQL-like filters.
The framework then answers the query based on these filters,
without requiring the user to make changes to the core implementation of the
framework or to reorganize the input data of the analysis.
The challenges that arise due to the amounts of data to be processed
and due to the complexity of the required computations are largely
encapsulated within the framework and hidden from the user. 

Our prototype implementation of this framework for Query-Driven Large-Scale
Portfolio Aggregate Risk Analysis, referred to as QuPARA, uses
Apache's Hadoop \cite{hadoop1, hadoop2} implementation of the MapReduce
programming model \cite{map1, map2, map3} to exploit parallelism, and Apache
Hive \cite{hiveql1, hiveql2} to support ad hoc queries. Even though QuPARA is
not as fast as a production system on the narrow set of standard portfolio
metrics, it can answer a wide variety of ad hoc queries in an efficient manner.
For example, our experiments demonstrate that an industry-size risk analysis
with 1,000,000 simulation trials, 1,000 events per trial, and on a portfolio
consisting of 3,200 risk transfer contracts (layers) with an average of 5 event
loss tables per layer can be carried out on a 16-node Hadoop cluster in just
over 20 minutes.

The remainder of this paper is organized as follows.
Section~\ref{sec:risk-analysis} gives an overview of reinsurance risk analysis. 
Section~\ref{sec:framework} proposes our new risk analysis framework.
Section~\ref{sec:queries} considers an example query processed by the framework
and various queries for fine-grained aggregate risk analysis.
Section~\ref{sec:implementation} describes implementation details.
Section~\ref{sec:experimentalstudies} presents a performance evaluation of our
framework.
Section~\ref{sec:conclusions} presents conclusions and discusses future work.

\section{An Overview of Risk Analysis}

\label{sec:risk-analysis}

A reinsurance company typically holds a \emph{portfolio} of programs that
insure primary insurance companies against large-scale losses, like those
associated with catastrophic events.
Each \emph{program} contains data that describes (1) the buildings to be insured
(the \emph{exposure}), (2) the modelled risk to that exposure (the \emph{event
loss tables}), and (3) a set of risk transfer contracts (the \emph{layers}).

The \emph{exposure} is represented by a table, one row per building covered,
that lists the building's location, construction details, primary insurance
coverage, and replacement value. The modelled risk is represented by an
\emph{event loss table} (ELT). This table lists for each of a large set of possible
catastrophic events the expected loss that would occur to the exposure should
the event occur. Finally, each \emph{layer} (risk transfer contract) is described by
a set of financial terms that includes aggregate deductibles and limits (i.e.,
deductibles and maximal payouts to be applied to the sum of losses over the year)
and per-occurrence deductibles and limits (i.e., deductibles and maximal payouts to
be applied to each loss in a year), plus other financial terms.

\begin{figure}
	\centering
	\includegraphics[width=0.35\textwidth]{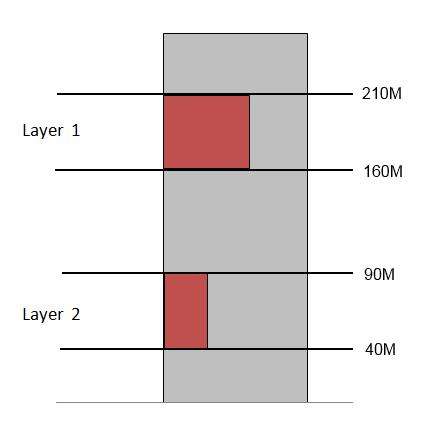}
	\caption{An example two-layer reinsurance program.}
	\label{fig_Layers}
\end{figure}

Consider, for example, a Japanese earthquake program. The exposure might list 2
million buildings (e.g., single-family homes, small commercial buildings, and
apartments) and, for each, its location (e.g., latitude and longitude),
constructions details (e.g., height, material, roof shape, etc.),
primary insurance terms (e.g., deductibles and limits), and replacement value.
The event loss table might, for
each of 100,000 possible earthquake events in Japan, give the sum of the losses
expected to the associated exposure should the associated event occur. Note
that ELTs are the output of stochastic region peril models \cite{catmodel-1} and 
typically also include some additional financial terms. Finally, a risk transfer
contract may consist of two layers as shown in
Figure~\ref{fig_Layers}. The first layer is a per-occurrence layer that pays
out a 60\% share of losses between 160 million and 210 million associated with
a single catastrophic event. The second layer is an aggregate layer
covering 30\% of losses between 40 million and 90 million that accumulate due
to earthquake activity over the course of a year.

Given a reinsurance company's portfolio described in terms of exposure, event
loss tables, and layers, the most fundamental type of
analysis query computes an Exceedance Probability (EP) curve, which
represents, for each of a set of user-specified loss values, the probability
that the total claims a reinsurer will have to pay out exceeds this value.
Not surprisingly there is no computationally feasible closed-form
expression for computing such an EP curve over hundreds of thousands of events
and millions of individual exposures. Consequently a simulation approach must
be taken. The idea is to perform a stochastic simulation based on a \emph{year
event table} (YET). This table describes a large number of trials, each
representing one possible sequence of catastrophic events that might occur
in a given year. This YET is generated by an
event simulator that uses the expected occurrence rate of each event plus other
hazard information like seasonality.
The process to generate the YET is beyond the scope of this paper, and we
focus on the computationally intensive task of computing the expected loss
distribution (i.e., EP curve)
for a given portfolio, given a particular YET.
Given the sequence of events in a given trial, the loss for this particular
trial can be computed, and the overall loss distribution is obtained from the
losses computed for the set of trials.

While computing the EP curve for the company's entire portfolio
is critical in assessing a company's solvency, analysts are often interested in
digging deeper into the data and posing a wide variety of queries with the goal
of analysing such things as cash flow throughout
the year, diversity of the portfolio, financial impact of adding a new contract
or contracts to the portfolio, and many others.

The following is a representative, but far from complete, set of example queries.
Note that while all of them involve some aspects of the basic aggregate risk
analysis algorithm used to compute Exceedance Probability curves, each puts
its own twist on the computation.

\begin{description}[leftmargin=12pt]
  \item[EP Curves with secondary uncertainty:]
    In the basic aggregate risk analysis algorithm, the loss value for each event
    is represented by a mean value. This is
    an oversimplification because for any given event there are a multitude of
    possible loss outcomes.
    This means that each event has an associated probability distribution of
    loss values rather than a single associate loss value.
    Secondary uncertainty arises from the fact that we are not just unsure
    whether an event will occur (\emph{primary uncertainty}) but also about
    many of the exposure and hazard parameters and their interactions.

    Performing aggregate risk analysis accounting for secondary uncertainty
    (i.e., working with the loss distributions associated with events rather
    than just mean values) is computationally intensive due to the statistical
    tools employed---for example, the beta probability distribution is employed
    in estimating the loss using the inverse beta cumulative density
    function \cite{ICCS2013}---but is essential in many applications.

  \item[Return period losses (RPL)]\textbf{by line of business (LOB), class
    of business (COB) or type of participation (TOP):} In the reinsurance
    industry, a layer defines coverage on different types of exposures and the
    type of participation.
    Exposures can be classified by class of business (COB) or line of business
    (LOB) (e.g., marine, property or engineering coverage).
    The way in which the contractual coverage participates when a catastrophic
    event occurs is defined by the type of participation (TOP). 
    Decision makers may want to know the loss distribution of a specific layer
    type in their portfolios, which requires the analysis to be restricted
    to layers covering a particular LOB, COB or TOP.

  \item[Region/peril losses:] This type of query calculates the expected losses or a loss
    distribution for a set of geographic regions (e.g., Florida or Japan),
    a set of perils (e.g., hurricane or earthquake) or a combination of
    region and peril.  This allows the reinsurer to understand both what types of
    catastrophes provide the most risk to their portfolio and in which regions of
    the globe they are most heavily exposed to these risks.
    This type of analysis helps the reinsurer to diversify or maintain a
    persistent portfolio in either dimension or both.

  \item[Multi-marginal analysis:] Given the current portfolio and a small set
    of potential new contracts, a reinsurer will have to decide which contracts
    to add to the portfolio.
    Adding a new contract means additional cash flow but also increases the
    exposure to risk.
    To help with the decision which contracts to add, multi-marginal analysis
    calculates the
    difference between the loss distributions for the current portfolio and
    for the portfolio with any subset of these new contracts added.
    This allows the insurer to choose contracts or to price the contracts so
    as to obtain the ``right'' combination of added cash flow and added risk.

  \item[Stochastic exceedance probability (STEP) analysis:] This
    analysis is a stochastic approach to the weighted convolution of multiple
    loss distributions.
    After the occurrence of a natural disaster not in their event catalogue,
    catastrophe modelling \cite{catmodel-1} vendors
    attempt to estimate the distribution of possible loss outcomes for that
    event. 
    One way of doing this is to find similar events in existing stochastic
    event catalogues and propose a weighted combination of the distributions of
    several events that best represents the actual occurrence.
    A simulation-based approach allows for the simplest method of producing
    this combined distribution.
    To perform this type of analysis, a customized Year Event Table must be
    produced from the selected events and their weights.
    In this YET, each trial contains only one event, chosen with a probability
    proportional to its weight. 
    The final result is a loss distribution of the event, including various
    statistics such as mean, variance and quantiles.
  \item[Periodic Loss Distribution:]
    Many natural catastrophes have a seasonal component to them, that is,
    do not occur uniformly throughout the year. For
    example, hurricanes on the Atlantic coast occur between July and
    November.
    Flooding in equatorial regions occurs in the rain season. As a result,
    the reinsurer may be interested in how their potential losses fluctuate
    throughout the year, for example to reduce their exposure through reduced
    contracts or increased retrocessional coverage during riskier periods.
    To aid in these decisions, a periodic loss distribution represents the
    loss distribution for different periods of the year, such as quarters,
    months, weeks, etc.

\end{description}

\section{QuPARA Framework}

\label{sec:framework}

In this section, we present our QuRARA framework.
Before describing QuPARA, we give an overview of the steps involved in
answering an aggregate query sequentially.
This will be helpful in understanding the parallel evaluation of aggregate
queries using QuPARA.


The loss distribution is computed from the portfolio and the YET in two
phases.
The first phase computes a \emph{year loss table} (YLT).
For each trial in the YET, each event in this trial, and each ELT that
includes this event, the YLT contains a tuple \tuple{trial, event, ELT, loss}
recording the loss incurred by this event, given the layer's financial terms
and the sequence of events in the trial up to the current event.
The second phase then aggregates the entries in the YLT to compute the
final loss distribution.
Algorithm~\ref{alg:sequential} shows the sequential computation of the YLT.

\begin{algorithm}
  \DontPrintSemicolon
  \caption{Sequential Aggregate Risk Analysis}
  \label{alg:sequential}
  \KwIn{Portfolio and YET}
  \KwOut{YLT}
  \BlankLine
  \For{each trial $T$ in the YET}{%
    \For{each event $X$ in $T$}{%
      \For{each program $P$ in the portfolio}{%
        \For{each layer $L$ in $P$}{%
          \For{each ELT $E$ covered by $L$}{
            Lookup $X$ in $E$ to determine the loss $l_X$ associated with $X$\;
            $l_{L} \leftarrow$ $l_{L}$ + $l_{X}$\;
          }
          Apply per-occurrence and aggregate financial terms to $l_{L}$\label{line:finterms}\;
          $l_{P} \leftarrow$ $l_{P}$ + $l_{L}$\;
        }
        $l_{PF} \leftarrow$ $l_{PF}$ + $l_{P}$\;
      }
      $l_{T} \leftarrow $ $l_{T}$ + $l_{PF}$\;
    }
    Populate $YLT$
  }
\end{algorithm}

\begin{figure*}
  \includegraphics[width=\textwidth]{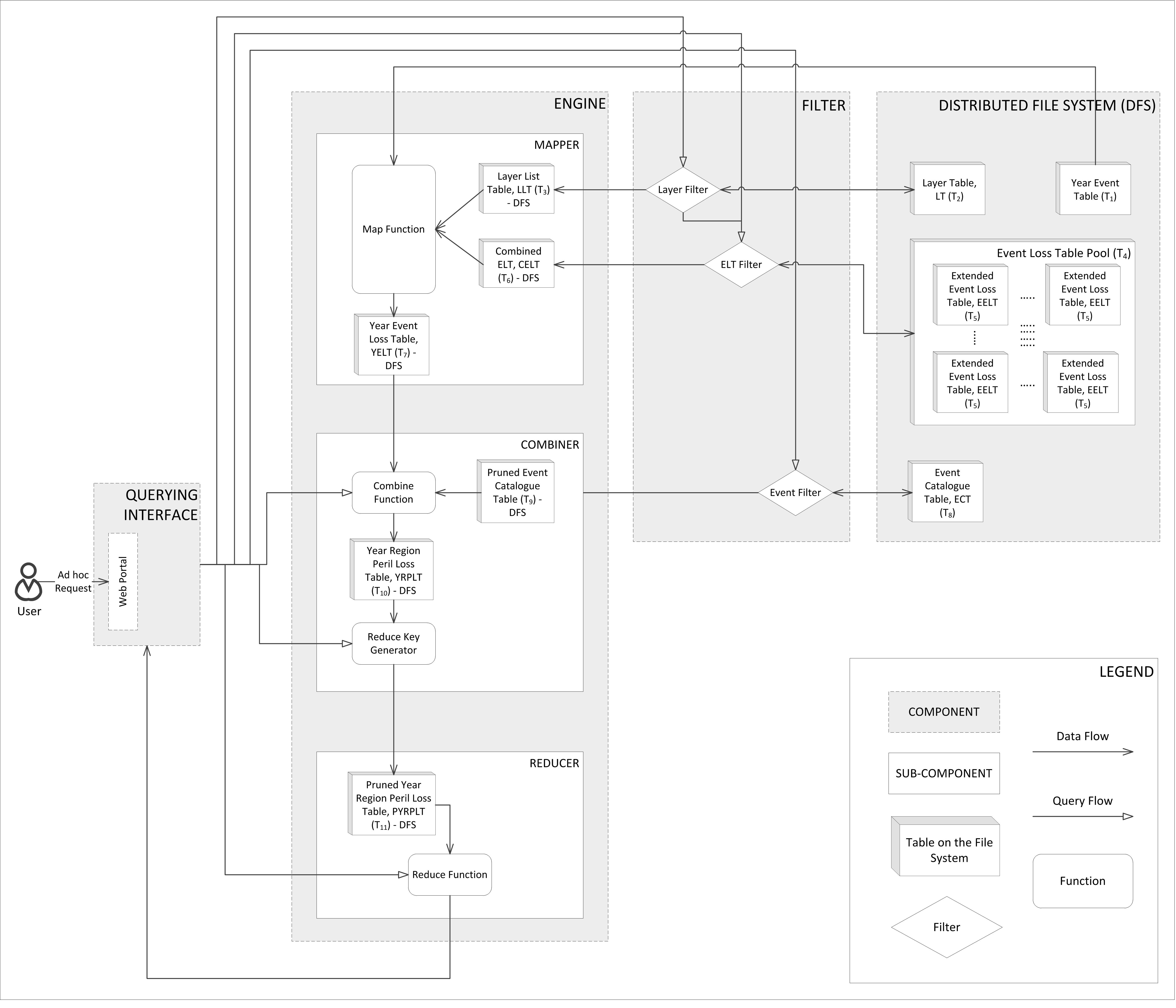}
  \caption{The Query-Driven Portfolio Aggregate Risk Analysis (QuPARA)
    Framework}
  \label{fig:architecture}
\end{figure*}

After looking up the losses incurred by a given event in the ELTs of a given
layer and summing these losses, the resulting \emph{layer loss} $l_L$ is reduced
by applying the layer's per-occurrence and aggregate financial terms.
The remaining loss is added to the \emph{program loss} $l_P$ for this event,
which in turn is added tho the \emph{portfolio loss} $l_{PF}$ for this event.
Finally, the loss of an entire trial, $l_T$ is computed by summing the
portfolio losses for all events in the trial.
Depending on the level of detail of the final analysis, the YLT is populated
with the loss values at different levels of aggregation.
At one extreme, if only the loss distribution for the entire portfolio is
of interest, there is one loss value per trial.
At the other extreme, if the filtering of losses based on some combination
of region, peril, LOB, etc. is required, the YLT contains one loss value
for each $\langle$trial, program, layer, ELT, event$\rangle$ tuple.

In order to answer ad hoc aggregate queries efficiently, our QuPARA framework
provides a parallel implementation of such queries using the MapReduce
programming model.
The computation of the YLT is the responsibility of the mapper, while
the computation of the final loss distribution(s) is done by the reducer.
Next we describe the components of QuPARA.

\subsection{Components}

Figure~\ref{fig:architecture} visualizes the design of QuPARA.
The framework is split into a front-end offering a \emph{query interface}
to the user, and a back-end consisting of a \emph{distributed file system}
and a \emph{query engine} that executes the query using MapReduce.
As already stated, the user poses a query to the query interface using an
SQL-like syntax and does not need to have any knowledge of the implementation
of the back-end.


The \emph{query interface} offers a web-based portal through which the user can
issue ad hoc queries in an SQL-like syntax.
The query is then parsed and passed to the query engine.

The \emph{distributed file system}, implemented using Hadoop's HDFS, stores
the raw portfolio data, as well as the YET in tabular form.

The \emph{query engine} employs the MapReduce framework to evaluate the
query using a single round consisting of a map/combine step and a reduce step.
During the map step, the engine uses one mapper per trial in the YET, in order
to construct a YLT from the YET, equivalent to the
sequential construction of the YLT from the YET using
Algorithm~\ref{alg:sequential}.
The combiner and reducer collaborate to aggregate the loss information in the
YLT into the final loss distribution for the query.
There is one combiner per mapper.
The combiner pre-aggregates the loss information produced by this mapper,
in order to reduce the amount of data to be sent across the network to the
reducer(s) during the shuffle step.
The reducer(s) then carry out the final aggregation.
In most queries, which require only a single loss distribution as output,
there is a single reducer.
Multi-marginal analysis is an example where multiple loss distributions are
to be computed, one per subset of the potential contracts to be added to
the portfolio.
In this case, we have one reducer for each such subset, and each reducer
produces the loss distribution for its corresponding subset of contracts.

Each mapper retrieves the set of ELTs required for the query from the
distributed file system using a layer filter and an ELT filter.
Specifically, the query may specify a subset of the layers in the portfolio
to be the subject of the analysis.
The layer filter retrieves the identifiers of the ELTs contained in these layers
from the layer table.
If the query specifies, for example, that the analysis should be restricted to
a particular type of peril, the ELT filter then extracts from this set of
ELTs the subset of ELTs corresponding to the specified type of peril.
Given this set of ELT identifiers, the mapper retrieves the actual ELTs and,
in memory, constructs a \emph{combined ELT} associating a loss with each
\tuple{event, ELT} pair.
It then iterates over the sequence of events in its trial, looks up the
ELTs recording non-zero losses for each event, and generates the corresponding
\tuple{trial, event, ELT, loss} tuple in the YLT, taking each ELT's financial
terms into account. 

The aggregation to be done by the combiner depends on the query.
In the simplest case, a single loss distribution for the selected set of layers
and ELTs is to be computed.
In this case, the combiner sums the loss values in the YLT output by the mapper
and sends this aggregate value to the reducer.
A more complicated example is the computation of a weekly loss distribution.
In this case, the combiner would aggregate the losses corresponding to the
events in each week and send each aggregate to a different reducer.
Each reducer is then responsible for computing the loss distribution for one
particular week.

A reducer, finally, receives one loss value per trial.
It sorts these loss values in increasing order and uses this sorted list
to generate a loss distribution. 

The following is a more detailed description of the mapper, combiner, and
reducer used to implement parallel aggregate risk analysis in MapReduce.

\subsubsection{Mapper}

The mapper, shown in Algorithm~\ref{alg:map}, takes as input an
entire trial and the list of its events, represented as a pair
\tuple{$T, E := \{E_1, E_2, \dots, E_m\}$}.
Algorithm~\ref{alg:map} does not show the construction of the combined ELT (CELT)
and layer list (LLT) performed by the mapper before carrying out the steps
in lines 1--9.
The loss estimate of an event in a portfolio is computed by scanning through
every layer $L$ in the LLT, retrieving and summing the loss estimates for
all ELTs covered by this layer, and finally applying the layer's financial terms.

\begin{algorithm}
  \caption{Mapper in parallel aggregate risk analysis}
  \label{alg:map}
  \DontPrintSemicolon
  \KwIn{\tuple{$T, E := \{E_1, E_2, \cdots , E_m\}$}, where $m$ is the
    number of events in a trial}
  \KwOut{A list of entries \tuple{$T, E_i, l_{PF}$} of the YLT}
  \BlankLine
  \For{each event, $E_{i}$ in $E$}{
    Look up $E_{i}$ in the CELT and find corresponding losses,
    $l_{E_i} = \{l^1_{E_i}, l^2_{E_i}, \cdots , l^n_{E_i}\}$, where 
    ELT$_1$, ELT$_2$, \dots, ELT$_n$ are the ELTS in the CELT\;
    \For{each layer, $L$, in the LLT}{
      \For{each ELT ELT$_{j}$ covered by $L$}{
        Lookup $l^j_{E_i}$ in $l_{E_i}$\;
        $l_{L} \leftarrow$ $l_{L}$ + $l^j_{E_i}$\;
      }
      Apply $L$'s financial terms to $l_{L}$\;
      $l_{PF} \leftarrow l_{PF} + l_{L}$
    }
    Emit(\tuple{$T$, $E_i$, $l_{PF}$})
  }
\end{algorithm}

\subsubsection{Combiner}

The combiner, shown in Algorithm~\ref{alg:combine}, receives as
input the list of triples \tuple{$T$, $E_i$, $l_{PF}$} generated by a single
mapper, that is, the list of loss values for the events in one specific trial.
The combiner groups these loss values according to user-specified grouping
criteria and outputs one aggregate loss value per group.

\begin{algorithm}
  \caption{Combiner in parallel aggregate risk analysis}
  \label{alg:combine}
  \DontPrintSemicolon
  \KwIn{A list of YLT entries \tuple{$T$, $E_i$, $l_{PF}$} for the events
  in a given trial $T$.}
  \KwOut{A list of aggregate YLT entries \tuple{$G_i$, $T$, $l_{G}$} with key
  $G_i$ for the event groups in trial $T$}
  \BlankLine
  Join input tuples with event catalogue to annotate events with their
  attributes (region, peril, etc.)\;
  Group events in the input list by event features according to the
  user's query\;
  \For{each group $G_{i}$}{
    $l_{G_i} \leftarrow{}$sum of the loss values associated with the events in
    $G_i$ in trial $T$\;
    Emit(\tuple{$T$, $G_{i}$, $L_{G}$})
  }
\end{algorithm}

\subsubsection{Reducer}

The reducer, shown in Algorithm~\ref{alg:reduce}, receives as
input the loss values for one specific group and for all trials in the YET.
The reducer then aggregates these loss values into the loss statistic
requested by the user.
For example, to generate an exceedance probability curve, the reducer
sorts the received loss values in increasing order and, for each loss value $v$
in a user-specified set of loss values, reports the percentage of trials with a 
loss value greater than $v$ as the probability of incurring a loss higher than
$v$. 

\begin{algorithm}
  \caption{Reducer in parallel aggregate risk analysis}
  \label{alg:reduce}
  \DontPrintSemicolon
  \KwIn{A list of loss tuples \tuple{$G_i$, $T$, $l_{PF}$} for an event
  group $G_i$.}
  \KwOut{Loss statistics for event group $G_i$ based on user's query}
  \BlankLine
  Based on user query, generate: \linebreak (i) Group loss distribution,
  or \linebreak(ii) Group loss statistics, or
  \linebreak(iii) Group value-at-risk (VaR) and/or Tail value-at-risk
  (TVaR), or \linebreak(iv) Exceedance probability curves\;
\end{algorithm}

\subsection{Data Organization}

The data used by QuPARA is stored in a number of tables: 
\begin{itemize}[leftmargin=\parindent]
  \item The year event table \texttt{YET} contains
    tuples \tuple{\texttt{trial\_ID, event\_ID, time\_Index, z\_PE}}.
    \texttt{trial\_ID} is a unique identifier associated with each of the
    one million trials in the simulation.
    \texttt{event\_ID} is a unique identifier associated with each event in
    the event catalogue.
    \texttt{time\_Index} determines the position of the occurrence of the event
    in the sequence of events in the trial.
    \texttt{z\_PE} is a random number specific to the program and event occurrence.
    Each event occurrence across different programs has a different associated
    random number. 
  \item The layer table \texttt{LT} contains tuples
    \tuple{\texttt{layer\_ID, cob, lob, tob, elt\_IDs}}.
    \texttt{layer\_ID} is a unique identifier associated with each layer in the
    portfolio.
    \texttt{cob} is an industry classification according to perils insured and
    the related exposure and groups homogeneous risks.
    \texttt{lob} defines a set of one or more related products or services
    where a business generates revenue.
    \texttt{tob} describes how reinsurance coverage and premium payments are
    calculated.
    \texttt{elt\_IDs} is a list of event loss table IDs that are covered by
    the layer.
  \item The layer list table \texttt{LLT} contains tuples
    \tuple{\texttt{layer\_ID, occ\_Ret, occ\_Lim, agg\_Ret, agg\_Lim}}. 
    Each entry is a simplified representation of the layer identified by
    \texttt{layer\_ID}.
    \texttt{occ\_Ret} is the occurrence retention or deductible of the insured
    for an individual occurrence loss.
    \texttt{occ\_Lim} is the occurrence limit or coverage the insurer will pay
    for occurrence losses in excess of the occurrence retention.
    \texttt{agg\_Ret} is the aggregate retention or deductible of the insured
    for an annual cumulative loss.
    \texttt{agg\_Lim} is the aggregate limit or coverage the insurer will pay
    for annual cumulative losses in excess of the aggregate retention.
  \item The event loss table pool \texttt{ELTP} contains tuples
    \tuple{\texttt{elt\_ID, region, peril}}.
    Each such entry associates a particular type of peril and a particular
    region with the ELT with ID \texttt{elt\_ID}.
  \item The extended event loss table \texttt{EELT} contains tuples
    \tuple{\texttt{event\_ID, z\_E, mean\_Loss, sigma\_I, sigma\_C, max\_Loss}}.
    \texttt{event\_ID} is the unique identifier of an event in the event catalogue.
    \texttt{z\_E} is a random number specific to the event occurrence.
    Event occurrences across different programs have the same random number.
    \texttt{mean\_Loss} denotes the mean loss incurred if the event occurs.
    \texttt{max\_Loss} is the maximum expected loss incurred if the event occurs.
    \texttt{sigma\_I} represents the variance of the loss distribution for this
    event.
    \texttt{sigma\_C} represents the error of the event-occurrence dependencies.
  \item The combined event loss table \texttt{CELT} is not stored on disk
    but is constructed by each mapper in memory from the extended event loss
    tables corresponding to the user's query.
    It associates with each event ID \texttt{event\_ID} a list of tuples
    \tuple{\texttt{elt\_ID, z\_E, mean\_Loss, sigma\_I, sigma\_C, max\_Loss}},
    which is the loss information for event \texttt{event\_ID} stored in
    the extended ELT \texttt{elt\_ID}.
  \item The year event loss table \texttt{YELT} is an intermediate table
    produced by the mapper for consumption by the combiner.
    It contains tuples
    \tuple{\texttt{trial\_ID, event\_ID, time\_Index, estimated\_Loss}}.
  \item The event catalogue \texttt{ECT} contains tuples
    \tuple{\texttt{event\_ID, region, peril}} associating a region
    and a type of peril with each event.
  \item The year region peril loss table \texttt{YRPLT} contains tuples
    \tuple{\texttt{trial\_ID, time\_Index, region, peril, estimated\_Loss}},
    listing for each trial the estimated loss at a given time, in a given
    region, and due to a particular type of peril.
    This is yet another intermediate table, which is produced by the combiner
    for consumption by the reducer.
\end{itemize}

\subsection{Data Filters}

QuPARA incorporates three types of data filters that allow the user to
focus their queries on specific geographic regions, types of peril, etc.
These filters select the appropriate entries from the data tables they
operate on, for further processing my the mapper, combiner, and reducer.

\begin{itemize}[leftmargin=\parindent]
  \item The \emph{layer filter} extracts the set of layers from the layer
    table \texttt{LT} and passes this list of layers to the mapper as the
    ``portfolio'' to be analyzed by the query.
    The list of selected layers is also passed to the ELT filter for selection
    of the relevant ELTs.
  \item The \emph{ELT filter} is used to select, from the ELT pool
    \texttt{ELTP}, the set of ELTs required by the
    layer filter and for building the combined ELT \texttt{CELT}.
  \item The \emph{event filter} selects the features of events from the event
    catalogue \texttt{ECT} to provide the grouping features to the combiner.
\end{itemize}

\section{An Example Query}

\label{sec:queries}

An example of an ad hoc request on QuPARA is to generate a report on seasonal
loss value-at-risk (VaR) with confidence level of 99\% due to hurricanes and
floods that affects all commercial properties in different locations in Florida.
The user poses the query through the query interface, which translates the
query into a set of five SQL-like queries to be processed by the query engine:

\begin{description}
  \item[$Q_{1}$:] The first part of processing any user query is the query
    to be processed by the layer filter.
    In this case, we are interested in all layers covering commercial
    properties, which translates into the following SQL query:
\begin{verbatim}
  SELECT * FROM LT
  WHERE lob IN commercial
\end{verbatim}
   \item[$Q_2$:] The second query is processed by the ELT filter to extract
     the ELTs relevant for the analysis.
     In this case, we are interested in all ELTs covered by layers returned
     by query $Q_1$ and which cover Florida (FL) as the region and hurricanes
     (HU) and floods (FLD) as perils:
\begin{verbatim}
  SELECT elt_ID FROM ELTP
  WHERE elt_ID IN Q1
  AND region IN FL
  AND peril IN HU, FLD
\end{verbatim}
  \item[$Q_{3}$:] This query is provided to the event filter for retrieving
    event features required for grouping estimated losses in the YELT.
\begin{verbatim}
  SELECT event_ID AND region FROM YELT
\end{verbatim}
  \item[$Q_{4}$:] This query is provided to the combiner for grouping
    all events in a trial based on their order of occurrence. For example, if
    there are 100 events equally distributed in a year and need to be grouped
    based on the four seasons in a year, then the estimated loss for each
    season is the sum of 25 events that occur in that season.
\begin{verbatim}
  SELECT trial_ID,
  SEASON_SUM(estimated_Loss)
  FROM YELT
  GROUP BY time_Index
\end{verbatim}
  \item[$Q_{5}$:] This query is provided to the reducer to define the
    final output of the user request.
    The seasonal loss Value-at-Risk (VaR) with 99\% confidence level is
    estimated. 
\begin{verbatim}
  SELECT VaR IN 0.01 FROM YRPLT
\end{verbatim}
\end{description}

\section{Implementation}

\label{sec:implementation}


For our implementation of QuPARA, we used Apache Hadoop,
an open-source software framework that implements the MapReduce programming model
\cite{hadoop1, hadoop2, hadoop3}.
We chose Hadoop because other available frameworks \cite{amazon, google}
require the use of additional interfaces, commercial or web-based, for deploying
an application. 

A number of interfaces provided by Hadoop, such as the \texttt{\small InputFormat} and 
the \texttt{\small OutputFormat} are implemented as classes. The Hadoop framework works 
in the following way for a MapReduce round.
The data files are stored on the Hadoop Distributed File System (HDFS) \cite{hadoop4}
and are loaded by the mapper.
Hadoop provides an \texttt{InputFormat} interface, which specifies how the
input data is to be split for processing by the mapper.
HDFS provides an additional functionality, called \emph{distributed cache}, for
distributing small data files that are shared by the nodes of the cluster.
The distributed cache provides local access to the shared data.
The \texttt{Mapper} interface receives the partitioned data and emits
intermediate key-value pairs.
The \texttt{Partitioner} interface receives the intermediate key-value pairs
and controls the partitioning of these keys for the \texttt{Reducer} interface.
Then the \texttt{Reducer} interface receives the partitioned intermediate
key-value pairs and generates the final output of this MapReduce round.
The output is received by the \texttt{OutputFormat} interface and provides it
back to HDFS. 

\begin{figure}
  \includegraphics[width=\columnwidth]{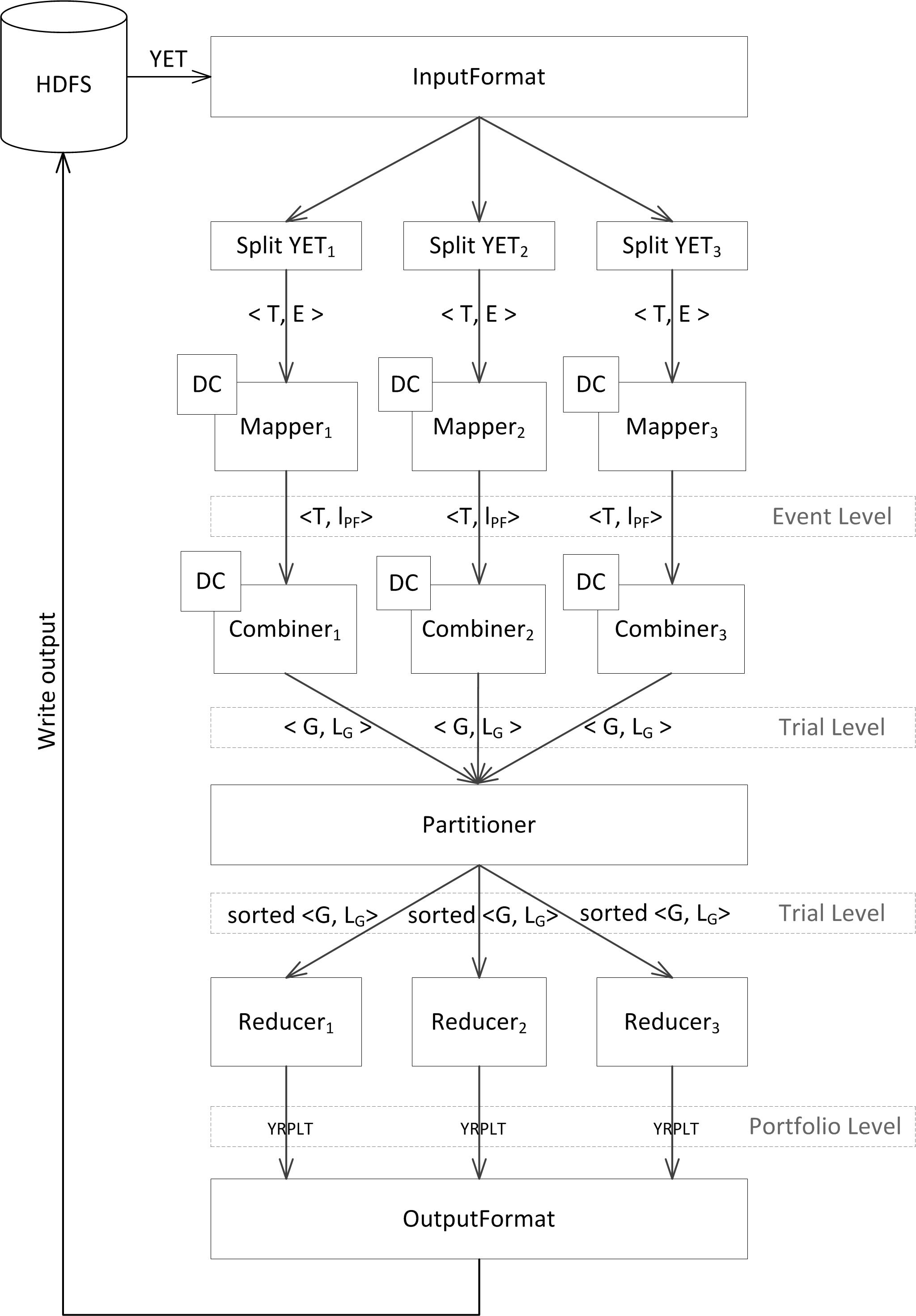}
  \caption{Apache Hadoop for MapReduce in the QuPARA Framework}
  \label{fig:MapReduce1}
\end{figure}

The input data for a MapReduce round in QuPARA is the year event table
\texttt{YET}, the event loss table pool \texttt{ELTP}, the list of layers \texttt{LT}, and the event
catalogue \texttt{ECT}, which are stored on HDFS.
The master node executes Algorithm~\ref{alg:masternode} and requires access to
the ELT pool and the portfolio 
before the MapReduce round. Firstly, the master node 
uses the portfolio to decompose the aggregate analysis job into a set of sub-jobs 
$\{J_{1}, \cdots , J_{x}\}$ as shown in line~1, each covering 200 layers.
This partition into sub-jobs was chose to ensure a balance between Input/Output
cost and the overhead for (sequentially) constructing the combined ELT in each
mapper; more sub-jobs means reading the YET more often, whereas fewer and larger
jobs increase the overhead of the sequential combined ELT construction.
The layers and the corresponding ELTs required by each sub-job are then
submitted to the Hadoop job scheduler in lines 2--6. 
The MapReduce round is illustrated in Figure \ref{fig:MapReduce1}.

\begin{algorithm}
  \caption{Algorithm for master node in aggregate risk analysis}
  \label{alg:masternode}
  \DontPrintSemicolon
  \KwIn{ELT pool, portfolio}
  \BlankLine
  Split the portfolio into jobs $J_{1}, \cdots , J_{x}$\;
  \For{each Job $J_{i}$}{
	Distribute its layers to nodes via the distributed cache\;
    \For{each layer $L_j$ in job $J_i$}{
    	Distribute the ELTs covered by $L_j$ to nodes via the distributed cache\;
  	}	
  Submit $J_{i}$ to Hadoop job scheduler\;
  }
\end{algorithm}

The \texttt{InputFormat} interface splits the YET based on the number of
mappers specified for the MapReduce round.
By default, Hadoop splits large files based on HDFS block size, but in this
paper, the \texttt{InputFormat} is redefined to split files based on the number
of mappers.
The mappers are configured so that they receive all the ELTs covered by the
entire portfolio via the distributed cache.
Each mapper then constructs its own copy of the combined ELT form the ELTs
in the distributed cache.
Using a combined ELT speeds up the lookup of events in the relevant ELTs
in the subsequent processing of the events in the YET.
If individual ELTs were employed, then one lookup would be required for fetching
the loss value of an event from each ELT.
Using a combined ELT, only a single lookup is required.
The mapper now implements Algorithm~\ref{alg:map} to compute the loss information
for all events in its assigned portion of the YET.

The combiner implements Algorithm \ref{alg:combine} to group the event-loss
pairs received from the mapper and emits the group-loss pairs to the partitioner.
The event catalogue is contained in the distributed cache of the combiner and
is used by the combiner to implement the grouping of events.
The Combiner delivers the grouped loss pairs to the partitioner.
The partitioner ensures that all the loss pairs with the same group
key are delivered to the same reducer.

The reducer implements Algorithm \ref{alg:reduce} to collect the sorted
group-loss pairs and produces year region peril loss table \texttt{YRPLT}.
Based on the query, the \texttt{OutputFormat} generates reports which are then
saved to the HDFS.

The layer filter, ELT filter, and event filter, described earlier, are implemented
using Apache Hive \cite{hiveql1}, which is built on top of the Hadoop Distributed File System
and supports data summarization and ad hoc queries using an SQL-like language.

\section{Performance Evaluation}

\label{sec:experimentalstudies}

In this section, we discuss our experimental setup for evaluating the
performance of QuPARA and the results we obtained.

\subsection{Platform}

We evaluated QuPARA on the Glooscap cluster of the
Atlantic Computational Excellence Network (ACEnet). We used
16 nodes of of the cluster, each of which was an SGI C2112-4G3 with four
quad-core AMD Opteron 8384 (2.7 GHz) processors and 64 GB RAM per node.
The nodes were connected via
Gigabit Ethernet. The operating system on each node was Red Hat Enterprise Linux
4.8. The global storage capacity of the cluster was 306 TB of Sun SAM-QFS, a
hierarchical storage system using RAID 5. Each node had 500GB of local storage.
The Java Virtual Machine (JVM) version was 1.6. The Apache Hadoop version
was 1.0.4. The Apache Hive version was 0.10.0.

\subsection{Results}

\begin{figure}
  \includegraphics[width=\columnwidth]{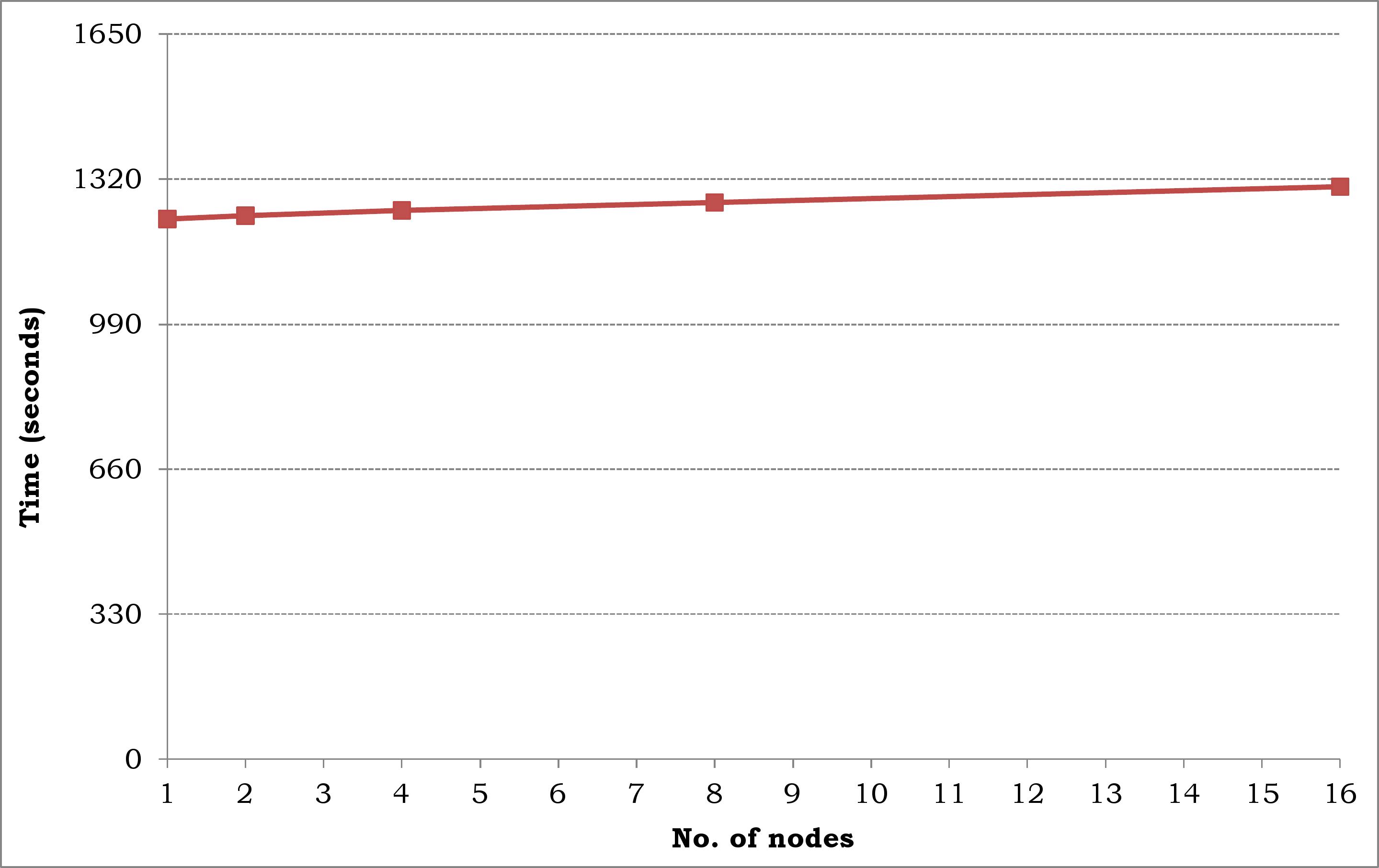}
  \caption{Running time of QuPARA on an increasing number of nodes with a fixed
  number of layers per job and one job per node}
  \label{fig:graph1}
\end{figure}

Figure \ref{fig:graph1} shows the total time taken in seconds for performing
aggregate risk analysis on our experimental platform.
Up to 16 nodes were used in the experiment.
Each node processed one job with 200 layers, each covering 5 unique ELTs.
Thus, up to 16,000 ELTs were considered.
The YET in our experiments contained 1,000,000 trials, each consisting of 1,000
events.
The graph shows a very slow increase in the total running time, in spite of
the constant amount of computation to be performed by each node (because
every node processes the same number of layers, ELTs, and YET entries).
The gradual increase in the running time is due to the increase in the setup
time required by the Hadoop job scheduler and the
increase in network traffic (reflected in an increase in the time taken by the
reducer).
Nevertheless, this scheduling and network traffic overhead amounted to only
1.67\%--7.13\% of the total computation time.
Overall, this experiment demonstrates that, if the hardware scales with the
input size, QuPARA's processing time of a query remains nearly constant.

\begin{figure}
  \includegraphics[width=\columnwidth]{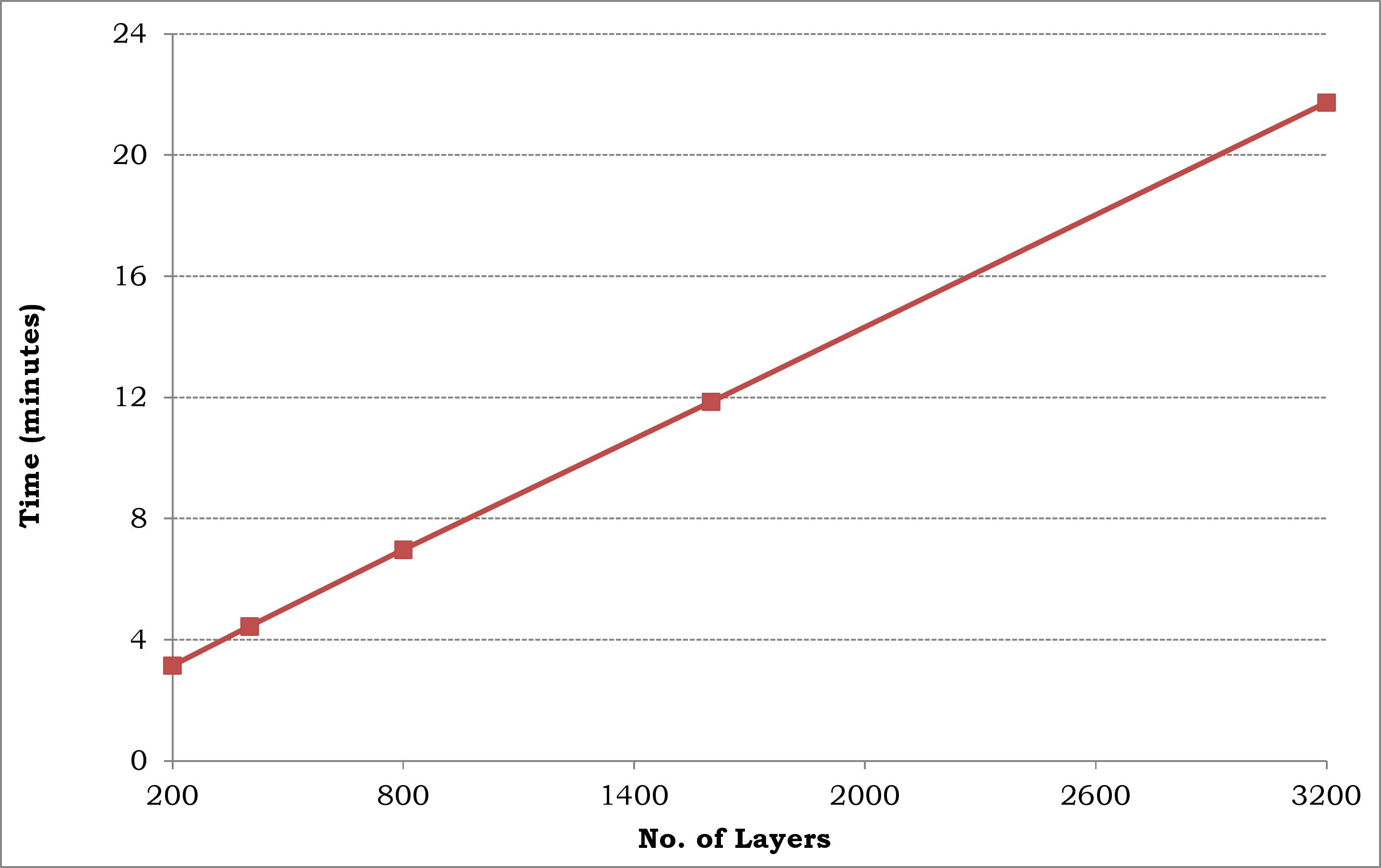}
  \caption{Running time of QuPARA on an increasing number of layers with a
  fixed number of nodes}
  \label{fig:graph2}
\end{figure}

Figure \ref{fig:graph2} shows the increase in running time when the number of
layers is increased from 200 to 3200 while keeping the number of nodes fixed.
Once again, each layer covered 5 ELTs, the YET contains 1,000,000 trials,
each consisting of 1,000 events.
16 nodes were used in this experiment.
As expected, the running time of QuPARA increases linearly with the input size
as we increase the number of layers.
With the increase in the number of layers, the
time taken for setup, I/O time, and the time for
all numerical computations scale in a linear fashion.
The time taken for building the combined ELT, by the reducer, and for clean up
are a constant. For 200 Layers, only 30\% of the time is taken for computations,
and the remaining time accounts for system and I/O overheads.
For 3200 layers, on the other hand, more than 80\% of the time is spent on
computations.

\begin{figure}
  \includegraphics[width=\columnwidth]{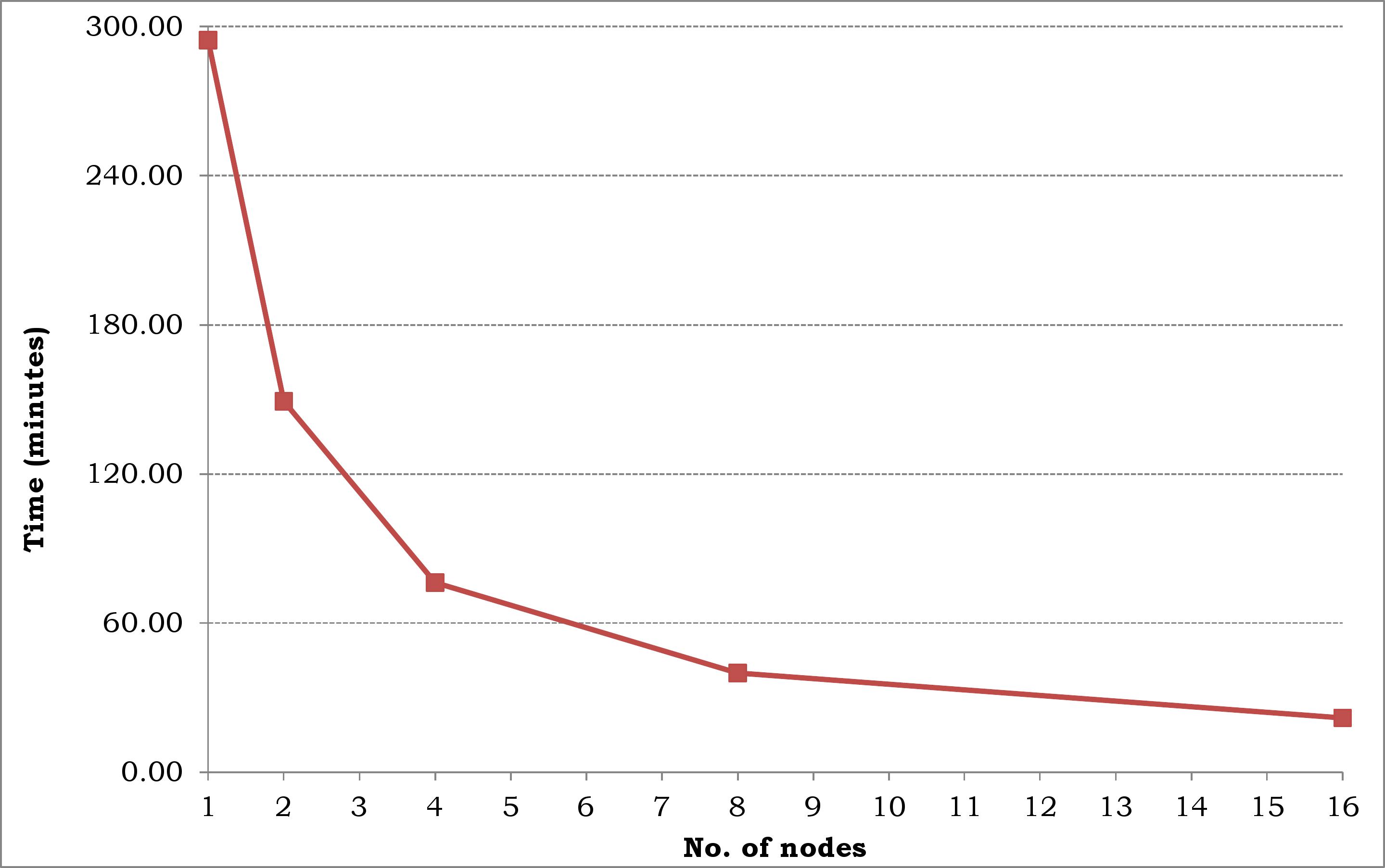}
  \caption{Running time of QuPARA on a fixed input size using between
  1 and 16 nodes}
  \label{fig:graph3}
\end{figure}

\begin{figure}
  \includegraphics[width=\columnwidth]{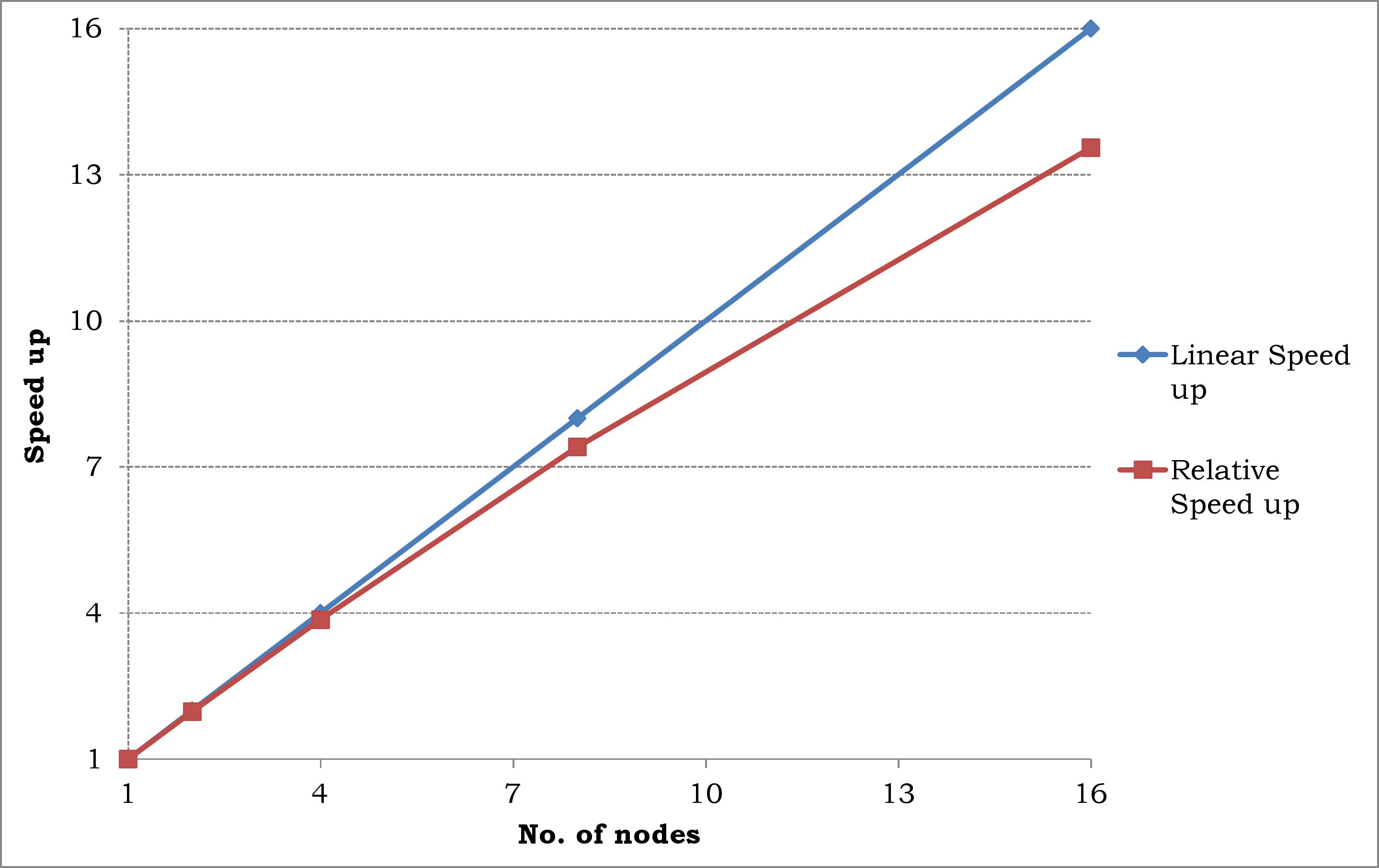}
  \caption{Speed-up achieved on the QuPARA framework}
  \label{fig:graph4}
\end{figure}

Figure \ref{fig:graph3} shows the decrease in running time when the number of
nodes is increased but the input size is kept fixed at 3,200 layers divided
into 16 jobs of 200 layers.
Figure \ref{fig:graph4} shows the relative speed-up achieved in this experiment,
that is, the ratio between the running time achieved on a single node and
the running time achieved on up to 16 nodes.
Up to 4 nodes, the speed-up is almost linear.
Beyond 4 nodes, the speed-up starts to decrease.
This is not due to an increase in overhead, but since the computation
times have significantly reduced after 4 nodes, the overheads start to take away
a greater percent of the total time.

In summary, our experiments show that QuPARA is capable of performing
ad-hoc aggregate risk analysis queries on industry-size data sets in a matter
of minutes.
Thus, it is a viable tool for analysts to carry out such analyses interactively.

\section{Conclusions}

\label{sec:conclusions}

Typical production systems performing aggregate risk analysis used in the industry 
are efficient for generating a small set of key portfolio metrics such as PML or TVAR 
required by rating agencies and regulatory bodies and essential for decision making. 
However, production systems can neither accommodate nor solve ad hoc queries that 
provide a better view of the multiple dimensions of risk that can impact a portfolio. 

The research presented in this paper proposes a framework for 
portfolio risk analysis capable of expressing and solving a variety of ad hoc catastrophic 
risk queries, based on MapReduce, and provided a prototype implementation, QuPARA,
using the Apache Hadoop implementation of the MapReduce programming model
and Apache Hive for expressing ad hoc queries in an SQL-like language.
Our experiments demonstrate the feasibility of answering ad hoc queries on industry-size
data sets efficiently using QuPARA.
As an example, a portfolio analysis on 3,200 layers and using a YET with 1,000,000
trials and 1,000 events per trial took less than 20 minutes.

Future work will aim to develop a distributed system by providing an online interface to 
QuPARA which can support multiple user queries. 





\end{document}